\documentclass[prd,twocolumn,showpacs]{revtex4}
\usepackage{graphicx}
\usepackage{latexsym}
\usepackage{amsmath}

 \newcommand{\bq}{\begin{equation}}
 \newcommand{\eq}{\end{equation}}
 \newcommand{\bqn}{\begin{eqnarray}}
 \newcommand{\eqn}{\end{eqnarray}}
 \newcommand{\nb}{\nonumber}
 \newcommand{\lb}{\label}

\begin{document}
\title{Cosmological constant and late transient acceleration of the universe
in the Horava-Witten Heterotic M-Theory on $S^{1}/Z_{2}$}
\author{Yungui Gong ${}^{1}$, Anzhong Wang ${}^{2,3}$ and Qiang Wu ${}^{2}$ }
\affiliation{${}^{1}$ College of
Mathematics $\&$ Physics, Chongqing University of Posts $\&$
Telecommunications, Chongqing 400065,   China\\
${}^{2}$ GCAP-CASPER, Physics Department, Baylor University,
Waco, TX 76798-7316\\
 ${}^{3}$ Department of Theoretical Physics, the State University of Rio de Janeiro, 
RJ, Brazil}
 
\begin{abstract}

Orbifold branes are studied in the framework of the 11-dimensional Horava-Witten 
heterotic M-Theory. It is found that the effective cosmological constant can be easily 
lowered to its current observational value by the mechanism of large extra 
dimensions.  The domination of this constant over the evolution of the universe 
is only temporarily.  Due to the interaction of the bulk and the branes, the 
universe will be in its decelerating expansion phase again in the future, 
whereby all problems connected with a far future de Sitter universe are resolved.

\end{abstract}
 
\maketitle

\section{Introduction}

Recent observations of supernova (SN) Ia  reveal 
the striking discovery that our universe has lately been in its accelerated 
expansion phase  \cite{agr98}. Cross checks from the cosmic microwave background 
radiation  and large scale structure all confirm this unexpected result \cite{Obs}. 
Such an expansion was predicted neither by the standard model of particle 
physics nor by the standard model of cosmology. In fact,  
in order to have an accelerated expansion, the latter requires
the introduction of either a tiny positive cosmological constant $\Lambda$ 
or an exotic component of matter that has a very large negative pressure 
and interacts with other components of matter weakly, if there is any. This 
invisible component is usually dubbed as dark energy. 

A tiny $\Lambda$ is well consistent with all observations carried out so far 
\cite{SCH07}, and the recent Hubble Space Telescope observations of the nearby galaxy 
groups Cen A/M83, M81/M82, and their vicinities, are even in favor of it
 \cite{Ch07}. Although the introduction of  $\Lambda$ 
may be the simplest resolution of the crisis, considerations of its origin lead
to other severe problems: (a) Its theoretical expectation values  
 exceed observational limits by $120$ orders of magnitude  \cite{wen}.  
(b) Its corresponding energy density   is comparable with that 
of matter only recently. Otherwise, galaxies would have not been formed. 
Considering the fact that the energy density of matter depends on time, 
one has to explain why only {\em now} the two are in the same order. 
(c) Once  $\Lambda$  dominates the evolution of the universe, it dominates forever.   
An eternally accelerating universe seems not consistent with
string/M-Theory, because it is endowed with a cosmological event horizon
that prevents the construction of a conventional S-matrix describing particle
interaction \cite{Fish}. Other problems with an asymptotical de Sitter universe 
in the future were explored in \cite{KS00}.  
 
In view of all the above,   dramatically different models  
have been proposed, including quintessence \cite{quit},  DGP branes \cite{DGP00}, 
and   the $f(R)$ models \cite{FR}. For details, see 
\cite{DE} and references therein. It is fair to say that so far no convincing model
has been constructed.

In this Letter, we study the cosmological constant problem and the late transient 
acceleration  of the universe in the framework of the Horava-Witten
heterotic M-Theory on $S^{1}/Z_{2}$ \cite{HW96}. In particular, using the 
Arkani-Hamed-Dimopoulos-Dvali (ADD) mechanism of large extra dimensions \cite{ADD98}, 
we show that the effective $\Lambda$ on each of the two branes can be easily 
lowered to its current observational value. The domination of this term  
 is only temporarily.  Due to the interaction of the bulk 
and the brane, the universe will be in its decelerating expansion phase again in the 
future, whereby all problems connected with a far future de Sitter universe are 
resolved. 

Before showing our above claims, we note that recently some attempts were made to 
derive a late time accelerating universe from string/M-Theory. In particular, 
Townsend and Wohlfarth \cite{townsend} invoked a time-dependent compactification  
of pure gravity in higher dimensions with hyperbolic internal space to circumvent 
Gibbons' non-go theorem \cite{gibbons}. Their exact solution   exhibits 
a short period of acceleration. The solution is the zero-flux limit of spacelike branes 
\cite{ohta}. If non-zero flux or forms are turned on,  a transient acceleration exists 
for both compact internal hyperbolic and flat spaces \cite{wohlfarth}. Other 
accelerating solutions  by compactifying more complicated time-dependent internal 
spaces can be found  in \cite{string}. In addition, in the string landscape \cite{Susk},
it is expected there are many different vacua with different local cosmological constants
\cite{BP00}. Using the anthropic principle, one may select the low energy vacuum in which
we can exist. However, many theorists still hope to explain the problem without invoking
the existence of ourselves in the universe.

In addition, in 4D spacetime there exists Weinberg's no-go theorem for the adjustment of 
the CC \cite{wen}. However, in higher dimensional spacetimes, the 4D vacuum energy on the 
brane does not necessarily give rise to an effective 4D CC. Instead, it may only curve the 
bulk, while leaving the brane still flat \cite{CEG01}, whereby Weinberg's no-go theorem
is evaded.

\section{Model in the Horava-Witten Heterotic M-Theory} 

The 11D spacetime of the Horava-Witten M-Theory 
  is described by the metric \cite{LOSW99},
\bq
\lb{2.1}
ds^{2}_{11} = V^{-2/3}\gamma_{ab} dx^{a}dx^{b} 
- V^{1/3}\Omega_{nm}dz^{n}dz^{m},
\eq
where $ds^{2}_{CY,6} \equiv \Omega_{nm}dz^{n}dz^{m}$ denotes the Calabi-Yau 3-fold, 
and $V$  the Calabi-Yau volume modulus that measures the deformation of the Calabi-Yau 
space, and depends only on  $x^{a}$, where $a = 0, 1, ... , 4$. 
Then, the 5D effective action of the Horava-Witten theory is given by 
\cite{LOSW99,Chen06}
\bqn
\lb{2.2}
S_{5} &=&  \frac{1}{2\kappa^{2}_{5}}\int_{M_{5}}{\sqrt{\gamma} \left(R[\gamma]
 - \frac{1}{2} \left(\nabla\phi\right)^{2}
+ 6\alpha^{2} e^{-2\phi}\right)} \nb\\
& & + \sum_{i= 1}^{2}{\epsilon_{i} \frac{6\alpha}{\kappa^{2}_{5}}\int_{M^{(i)}_{4}} 
 {\sqrt{-g^{(i)}} 
 e^{-\phi}}},
\eqn
where $\epsilon_{1} = - \epsilon_{2} = 1$, and
\bq
\lb{2.2a}
\phi   \equiv \ln({V}),\;\;\;
\kappa^{2}_{5} \equiv \frac{\kappa^{2}_{11}}{v_{CY,6}}, 
\eq
with $v_{CY,6}$ being the volume of the Calabi-Yau space, 
\bq
\lb{2.2b}
v_{CY,6} \equiv \int_{X}{\sqrt{\Omega}}.
\eq
The constant $\alpha$ is related to the internal four-form that 
has to be included in the dimensional reduction. This four-form results from the source
terms in the 11D Bianchi identity, which are usually non-zero. 
$g^{(i)}$ 's are the reduced metrics on the two boundaries $M^{(i)}_{4}\; (i =1,2)$.  
It should be noted that in general the dimensional reduction 
of the graviton and the four-form flux generates a large number of fields. 
However, it is consistent to set all the fields zero except for the 5D 
graviton and the volume modulus. This setup implies that all components of the four-form 
now point in the Calabi-Yau directions. In addition, it can be shown that
the above action is indeed the bosonic sector of a minimal $N =1$ gauged supergravity
theory in 5D spacetimes coupled to chiral boundary theories.

To study cosmological models in the above setup, we add matter fields on each of the two 
branes, 
\bq
\lb{2.3}
S_{(i)}^{m} = - \int_{M^{(i)}_{4}}{\sqrt{-g^{(i)}} {\cal{L}}_{m}^{(i)}\left(\phi, \chi\right)},
\eq
where $\chi$ collectively denotes the SM fields localized on the branes. Clearly,
this in general makes the two branes no longer supersymmetric, although the bulk still is. 
Variation of the action,
\bq
\lb{2.3a}
S^{total}_{5} = S_{5} +  \sum_{i=1}^{2}{S_{(i)}^{m}},
\eq
with respect to $\gamma_{ab}$ yields the field equations, 
\bqn
\lb{2.3b}
{}^{(5)}G_{ab} &=&  \kappa^{2}_{5} {}^{(5)}T_{ab}^{\phi}  \nb\\
& & + \kappa^{2}_{5} \sum^{2}_{i =1}{{\cal{T}}^{(i)}_{\mu\nu}}e^{(\mu)}_{a}e^{(\nu)}_{b} 
\sqrt{-\frac{g^{(i)}}{\gamma}}\;\delta\left(\Phi_{i}\right),
\eqn
where 
$T_{ab}^{\phi}$ and ${\cal{S}}^{(i)}_{\mu\nu}$'s are the energy-momentum tensors of the bulk and
branes, respectively, and are given by 
\bqn
\lb{2.3c}
\kappa^{2}_{5} {}^{(5)}T_{ab}^{\phi} &\equiv& \frac{1}{2}
\left(\nabla _{a}\phi\right)\left(\nabla_{b}\phi\right)\nb\\
& &  - \frac{1}{4}\gamma_{ab}\left[\left(\nabla\phi\right)^{2} 
- 12 \alpha^{2}e^{-2\phi}\right],\nb\\
{\cal{T}}^{(i)}_{\mu\nu} &\equiv& \frac{6 \alpha\epsilon_{i}}{\kappa^{2}_{5}}
e^{-\phi} g_{\mu\nu}^{(i)} +   {\cal{S}}^{(i)}_{\mu\nu},\\
\lb{2.3d}
{\cal{S}}^{(i)}_{\mu\nu} &\equiv& 2 \frac{\delta {\cal{L}}_{m}^{(i)}}{\delta g^{(i)\; \mu\nu}}
-  g^{(i)}_{\mu\nu} {\cal{L}}_{m}^{(i)},\\ 
\lb{2.3e}
e^{a}_{(\mu)} &\equiv& \frac{\partial x^{a}}{\partial \xi^{\mu}},\nb\\
g^{(i)}_{\mu\nu} &\equiv & \left. e^{a}_{(\mu)}e^{b}_{(\nu)}\gamma_{ab}\right|_{M^{(i)}},
\eqn
where $\xi^{\mu} \; (\mu = 0, 1, 2, 3)$ are the intrinsic coordinates on the orbifold branes.
$\delta\left(\Phi_{i}\right)$ denotes the Dirac delta function, and the two orbifold branes are 
located on the hypersurfaces,
\bq
\lb{2.3f}
\Phi_{i}\left(x^{a}\right) = 0, \; (i = 1, 2).
\eq
It is interesting to note that the contribution of the modulus field to the branes acts as
a varying cosmological constant, as can be seen clearly from Eq.(\ref{2.3b}).    

Variation of the total action (\ref{2.3a}) with respect to $\phi$, on the other hand, 
yields the generalized Klein-Gordon equation,
\bqn
\lb{2.3g}
\Box{\phi} &=& 12 \alpha^{2} e^{-2\phi} + \sum^{2}_{i =1}{
\left({12 \alpha\epsilon_{i}}e^{-\phi} + \sigma^{(i)}_{\phi}\right)} \nb\\
& &   \times \sqrt{-\frac{g^{(i)}}{\gamma}}\;\delta\left(\Phi_{i}\right),
\eqn
where $\Box \equiv \gamma^{ab}\nabla_{a}\nabla_{b}$, and
\bq
\lb{2.3h}
\sigma^{(i)}_{\phi} \equiv  2 \kappa^{2}_{5}  \; \frac{\delta {\cal{L}}_{m}^{(i)}}{\delta \phi}.
\eq

\subsection{Spacetime in the Bulk}

Then,  it can be shown that the 5D  field equations in between the two orbifold branes
admit the solution,
\bqn
\lb{2.4}
ds_{5}^{2} &\equiv& \gamma_{ab} dx^{a}dx^{b} \nb\\
&=&  dt^{2} - \left(\frac{6}{5}\right)t^{2}\left(dy^{2} + \sinh^{2}y d\Omega_{3}^{2}\right),
\eqn
where $d\Omega_{3}^{2}$ is the metric on the unit 3-sphere.  The corresponding 
volume modulus is given by 
\bq
\lb{2.4a}
\phi = \ln\left(2\alpha t\right).
\eq
It should be noted that without 
the introduction of the matter fields (\ref{2.3}), this solution does not satisfy the reduced 
 field equations on the two branes \cite{Ben99}.  
From the expression of $ds^{2}_{5}$  we can see that the 5D spacetime is singular at $t = 0$, which
divides the whole manifold into two disconnected branches, $t > 0$ and $t < 0$. The branch 
$t < 0$ represents a collapsing spacetime starting from $t = - \infty$, while the one $t > 0$ 
represents an expanding universe starting from a big bang singularity at $t = 0$. Since in this 
Letter we are mainly interested in cosmological model, from now on we shall work only 
with the branch $t > 0$. Lifting the above solution to 11 dimensions, from Eqs.(\ref{2.2a}),
(\ref{2.4}) and (\ref{2.4a}) we find that the metric (\ref{2.1}) can be cast in the form,  
\bqn
\lb{2.5}
ds_{11}^{2} &=& \frac{1}{(2\alpha)^{2/3}}\left(d\tilde{t}^{2} - \frac{8}{15}\tilde{t}^{2}
\left(dy^{2} + \sinh^{2}y d\Omega_{3}^{2}\right)\right)\nb\\
& & - \left(2\alpha\right)^{1/3}\left(\frac{2\tilde{t}}{3}\right)^{1/2}ds_{CY,6}^{2},
\eqn
where $t = (2\tilde{t}/3)^{3/2}$. Clearly, the 11D spacetime is also singular at
$\tilde{t} = 0$, where the length of each of the 10 spatial dimensions shrinks to zero.
Such a singularity may be removed using the ideas from the resolution of curvature 
singularities in Loop Quantum Gravity \cite{RMS07}. However, since in this Letter we are mainly 
interested in the evolution of the universe in the late times, we shall not consider
this possibility here.   

\subsection{The  $S^{1}/Z_{2}$ Compactification}

To write down the field equations on the branes, one can first
express the delta function parts of ${}^{(5)}G_{ab}$ in terms of the discontinuities of
the first derivatives of the metric coefficients, and then equal the delta function parts 
of the two sides of Eq.(\ref{2.3b}), as shown systematically in \cite{WCS07}.
The other way is to use the Gauss-Codacci equations to write the $4$-dimensional 
Einstein tensor as  \cite{SMS00},
\bqn
\lb{2.6}
{}^{(4)}G_{\mu\nu} &=& \frac{2}{3} {}^{(5)}G_{ab} e^{a}_{(\mu)} e^{b}_{(\nu)} 
+ {}^{(5)}E_{\mu\nu}\nb\\
& & - \frac{2}{3}\left({}^{(5)}G_{ab}n^{a}n^{b} 
+ \frac{1}{4} {}^{(5)}G\right)g_{\mu\nu}\nb\\
  & & 
  + \left(K_{\mu\lambda}K^{\lambda}_{\nu} - K K_{\mu\nu}\right) \nb\\
  & & 
  - \frac{1}{2}g_{\mu\nu}\left(K_{\lambda\sigma}K^{\lambda\sigma}  - K^{2}\right),
  \eqn
where  
${}^{(5)}E_{\mu\nu}$ is the projection of the 5D Weyl tensor of the bulk onto the 
brane, defined as ${}^{(5)}E_{\mu\nu} \equiv {}^{(5)}C_{abcd} n^{a} 
e^{b}_{(\mu)}n^{c} 
e^{d}_{(\nu)}$, with $n^{a}$ being the normal vector to the brane. 
The extrinsic curvature $K_{\mu\nu}$ is defined as $K_{\mu\nu} 
\equiv e^{a}_{(\mu)} e^{b}_{(\nu)} \nabla_{a}n_{b}$, and 
${}^{(5)}G \equiv {}^{(5)}G_{ab}\gamma^{ab}$. 

Assuming that the two branes are located on the surfaces, $t = t_{i}(\tau_{i})$ and
$y = y_{i}(\tau_{i})$, we find that 
the normal vectors to the two branes are given by $n_{A}^{(i)} = 
e_{i} \sqrt{6/5} t_{i}\left(-\dot{y}_{i}\delta^{t}_{A} + \dot{t}_{i}\delta^{y}_{A}\right)$,
where $e_{i} = \pm 1$,  and $\tau_{i}$'s denote the proper times of the branes, 
defined by $d\tau_{i} = \sqrt{1 - \frac{6}{5}{t}_{i}^{2}(\dot{y}_{i}/\dot{t}_{i})^{2}} dt_{i}$. 
An overdot denotes the ordinary differentiation with respect to $\tau_{i}$.
When $e_{i} =  1$, the normal vector points in the $y$-increasing direction, and 
when $e_{i} =  -1$ it points in the $y$-decreasing direction.  
The reduced metrics on the two branes take
the form
\bq
\lb{2.7a}
\left. ds^{2}_{5}\right|_{M^{(i)}_{4}} \equiv g^{(i)}_{\mu\nu} d\xi^{\mu}_{(i)} d\xi^{\nu}_{(i)} 
= d\tau_{i}^{2} - a_{i}^{2}(\tau_{i})d\Omega_{3}^{2},
\eq
where $a_{i} \equiv \sqrt{6/5} \; t_{i}\sinh(y_{i})$.
 Assume that on each of the two branes there is a perfect fluid, ${\cal{S}}^{(i)}_{\mu\nu}
 = \tau^{(i)}_{\mu\nu} + \lambda^{(i)} g^{(i)}_{\mu\nu}$, where $\lambda^{(i)}$ is
 the cosmological constant on the ith brane, and 
$\tau^{(i)}_{\mu\nu} = \left(\rho^{(i)} + p^{(i)}\right)u^{(i)}_{\mu}u^{(i)}_{\nu} 
- p^{(i)} g^{(i)}_{\mu\nu}$, with $u^{(i)}_{\mu}  = \delta_{\mu}^{\tau_{i}}$. Then, 
from the Lanzos equations \cite{Lanzos},
\bq
\lb{2.7}
\left[K_{\mu\nu}\right]^{-} - g_{\mu\nu} \left[K\right]^{-} = - \kappa^{2}_{5}{\cal{T}}_{\mu\nu},
\eq
where $\left[K_{\mu\nu}\right]^{-} \equiv K_{\mu\nu}^{+} - K_{\mu\nu}^{-}, \; \left[K\right]^{-} \equiv
g^{\mu\nu} \left[K_{\mu\nu}\right]^{-}$, we obtain \cite{GW07},
\bqn
\lb{2.8a}
& & H^{2} + \frac{1}{a^{2}} = \frac{8\pi G}{3} \rho  +  \frac{\Lambda}{3} 
+ \frac{\kappa^{4}_{5}}{36}   \rho^{2} +  
\frac{\sinh^{2}y}{5 a^{2}} \nb\\
& & + 2\epsilon_{i}\left(\frac{\pi G}{5\rho_{\Lambda}}\right)^{1/2}
\left(\rho + 2\rho_{\Lambda}\right) \frac{\sinh y}{a},\\
\lb{2.8b}
& & \dot{\rho} + 3H(\rho + p) = - \frac{1}{\Delta}  
\left\{\frac{2}{5}H\dot{y}^{2} + \frac{\sinh{2}y}{5 a^{2}} \dot{y}\right.\nb\\
& & \left. + 2\epsilon_{i}\left(\frac{\pi G}{5\rho_{\Lambda}}\right)^{1/2}
\left(\rho + 2\rho_{\Lambda}\right)\left(\dot{y}\coth{y} - H\right)\frac{\sinh y}{a}\right\},\nb\\ 
\eqn
where 
$$
\Delta = \frac{4\pi G}{3}\left(2 + \frac{\rho}{\rho_{\Lambda}}\right)
+ 2\epsilon_{i}\left(\frac{\pi G}{5\rho_{\Lambda}}\right)^{1/2}
  \frac{\sinh y}{a}, 
$$
$\rho_{\Lambda} \equiv \Lambda/8\pi G, \; H \equiv \dot{a}/a$, and 
$G$ and $\Lambda$ are, respectively, the 4D Newtonian and effective 
cosmological constant, given by
\bq
\lb{2.9}
8\pi G \equiv \frac{1}{6}\kappa^{4}_{5} \lambda,\;\;\;\;
\Lambda \equiv \frac{1}{12}\left(\frac{48\pi G}{\kappa_{5}^{2}}\right)^{2}.
\eq
In writing the above expressions we had used the $Z_{2}$ symmetry, $K_{\mu\nu}^{+} = - K_{\mu\nu}^{-}$. For 
the sake of simplicity, we also dropped the indices ``i", without causing any confusions. 
It is interesting to note that ${}^{(4)}G_{\mu\nu}$ given by Eq.(\ref{2.6}) depends on 
$K_{\mu\nu}$ quadratically,
so that it does not depend on the signs of $e_{i}$, nor on the choice whether $K_{\mu\nu}^{+}$ or
$K_{\mu\nu}^{-}$ is going to be used. 

The third term  in the right-hand side of Eq.(\ref{2.8a}) represents the brane corrections,
which is important in the early epoch of the  evolution of the universe. The fourth term 
is the projection of the energy-momentum tensor of the scalar field onto the brane, while the
last term represents   the interaction between the brane and the bulk. It can be very important
in the late evolution of the universe, as to be shown below.  The first two terms  are those that 
also appear in Einstein's theory of gravity, although their physical origins are completely different. In
particular, in Einstein's theory   $\lambda$ is an arbitrary constant, while
here it is completely fixed by the  Newtonian constant $G$. In other words, it is now a fundamental
constant and plays the same role as $G$ does. On the other hand, assuming
that the typical size of the Calabi-Yau space is $R$, we find $v_{CY,6} \sim R^{6}$. Then,
from Eqs. (\ref{2.2a}), (\ref{2.9}) and the relation $\kappa^{2}_{D} = {M_{D}}^{2-D}$, 
we obtain
\bq
\lb{2.10}
\rho_{\Lambda} 
= 3\left(\frac{R}{l_{pl}}\right)^{12}
\left(\frac{M_{11}}{M_{pl}}\right)^{18}{M_{pl}}^{4},
\eq
where $M_{pl} \sim 10^{19} \; GeV$ and $l_{pl} \sim 10^{-35}\; m$ denote, respectively, the Planck mass 
and length. From the above expression we can see that, if the theory is in the TeV scale \cite{Anto00}, 
to have $\rho_{\Lambda} \simeq \rho_{ob}$ the typical size of the Calabi-Yau space $R$ needs to be 
only at the scale $10^{-22} \; m$, which is by far below the current observational constraints \cite{Hoyle}. 
If $M_{11} \sim 100\;TeV$ it needs to be at the scale $10^{-24} \; m$; 
and if $M_{11} \sim 10^{12}\; GeV$, it is at the Planck scale. Therefore, 
the Horava-Witten theory on $S^{1}/Z_{2}$ provides  a very viable mechanism to get 
$\rho_{\Lambda}$ down to its current observational value. Hence, the ADD mechanism that was initially 
designed to solve  the hierarchy problem \cite{ADD98} also solves the cosmological constant problem
in the Horava-Witten M-theory on $S^{1}/Z_{2}$. 

Note that to close the system of Eqs.(\ref{2.8a}) and (\ref{2.8b}), two additional equations are needed.
One of them can be the equation of the state of the perfect fluid, and the other is the equation that
describes the motion of the branes, given by
\bqn
\lb{2.11}
\dot{y} &=& \frac{5}{5\coth^{2}y -6}\left\{H\coth{y}\right.\nb\\
& &\left. 
+ \epsilon\left[\frac{6}{5}\left(H^{2} 
+ \frac{1}{a^{2}} - \frac{\sinh^{2}y}{5a^{2}}\right)\right]^{1/2}\right\},
\eqn
where $\epsilon = \pm 1$. From these equations it can be seen that by properly choosing the initial
positions of the branes $y_{i}(0)$, the last two terms in the right-hand side  
of Eq.(\ref{2.8a}) are neglected until very recently.
As a result, the evolution of the universe follows  almost the same trajectory as that described by
the standard $\Lambda$CDM model, in which the $\Lambda$ term dominates currently. However, the interaction 
between the bulk and the branes can be very important in the future, so that it leads to a late transient
acceleration of the universe. 

To show that this is exactly the case, we first fit our model with
observational data, and then use these best fitting data as our initial conditions to study the
future evolution of the universe.  Let us first parameterize the current density of each
component as $\Omega_m=\rho_0/\rho_{cr}$, $\Omega_\Lambda=\rho_\Lambda/\rho_{cr}$, and
$\Omega_k=3/(8\pi Ga_0^2\rho_{cr})$, where $\rho_{cr}=3H^2_0/(8\pi G)$. We use the 182 gold SN Ia 
data \cite{riess06} combined with Baryon Acoustic Oscillation (BAO) parameter from SDSS data \cite{sdss}. 
We find that the results depend on the choice of $\epsilon$ and $\epsilon_i$, where for the positive
(negative) tension brane we have $\epsilon_i = 1$ ($\epsilon_i = -1$), as can be seen from 
Eq.(\ref{2.2}).   In particular,  
for $\epsilon = 1 = \epsilon_i$, we find that the minimum of the function $\chi^2$ is $\chi^2=156.20$ and
$\Omega_m=0.28^{+0.20}_{-0.04}$, $\Omega_\Lambda=0.87^{+0.55}_{-0.15}$,
and $\Omega_k=0.16^{+0.00}_{-0.16}$. For $\epsilon=1 = - \epsilon_i$, 
we find $\chi^2=156.22$, $\Omega_m=0.22^{+0.09}_{-0.06}$, $\Omega_\Lambda=0.74^{+0.23}_{-0.05}$,
and $\Omega_k=0.08^{+0.00}_{-0.08}$. For $\epsilon=-1 = - \epsilon_i$, we find $\chi^2=156.20$, 
$\Omega_m=0.28^{+0.24}_{-0.04}$, $\Omega_\Lambda=0.87^{+0.41}_{-0.04}$,
and $\Omega_k=0.16^{+0.00}_{-0.16}$. And for $\epsilon=-1 = \epsilon_i$, we find $\chi^2=155.77$,
$\Omega_m=0.16^{+0.15}_{-0.07}$, $\Omega_\Lambda=0.71^{+1.12}_{-0.00}$, 
and $\Omega_k=0.09^{+0.00}_{-0.08}$. Taking these best fitting data as initial conditions, the future 
evolution of the acceleration of the universe is shown in Fig. 1, from which we can see that for 
the cases where $\epsilon = \epsilon_{i} = 1$ and $\epsilon = - \epsilon_{i} = 1$, after a finite 
time, the universe will become decelerating again. It should be noted that in all of these four cases 
$\rho_{m}$ always approaches to a very small value.  We set up a lowest limit for it, and once
 it reaches that limit, it will be set to zero for the rest of its evolution.

\begin{figure}
\centering
\includegraphics[width=8cm]{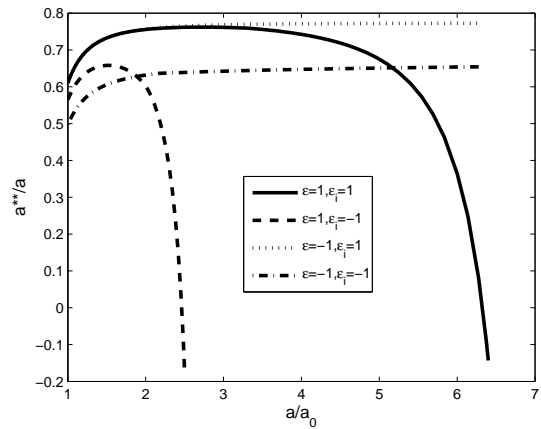}
\caption{The evolution of the quantity $a^{**}/a \equiv \ddot{a}/(a H^{2}_{0})$.   }
\label{fig1}
\end{figure}
 
\section{Main Results and Concluding Remarks} 

In this Letter, we have studied the evolution of the universe in the 
the framework of the 11-dimensional Horava-Witten  M-Theory on $S^{1}/Z_{2}$. We have shown 
explicitly  that the effective cosmological constant can be easily lowered to its current 
observational value using the large extra dimensions. The domination of this constant over 
the evolution of the universe is only currently.  Due to the interaction of the bulk and the 
branes, the universe will be in its decelerating expansion phase again in the future,
whereby all problems connected with a far future de Sitter universe are resolved.

It should be noted that  the expression of Eq.(\ref{2.10}) is quite general, and does not 
depend on the specific model considered in this Letter \cite{GW07}. 

It is also interesting to note that the mechanism used in this Letter is different from the 
so-called ``self-tuning" mechanism, proposed in both 5-dimensional spacetimes \cite{5CC} and 
in 6-dimensional spacetimes \cite{6CC}. In the 5D case, it was shown that hidden fine-tunings 
are required \cite{For00}, while in the 6D case it is still not clear whether loop corrections 
can be as small as  required by solving the CC problem \cite{Burg07}.

Two important problems have not been addressed in this Letter. One is the stability of radion
and the other is the constraints from observations. The former has been extensively
studied  in 5-dimensional spacetimes  \cite{branes}. The generalization of such studies to our 
model is straightforward, and is currently under our considerations. We are also investigating 
the constraints coming from the solar system tests \cite{solar}.
 
{\em Acknowledgments}: YGG is supported 
by   NSFC under grant No. 10447008 and 10605042, SRF for ROCS, 
State Education Ministry and CMEC under grant No. KJ060502.  
AW is partially supported by  NSFC under grant No. 10775119 and
 a VPR fund, Baylor University.

\end{document}